\documentclass[a4paper]{article}

\usepackage{INTERSPEECH2022}
\usepackage{amsmath,graphicx}
\usepackage{todonotes}
\usepackage{acronym}
\usepackage{multirow}
\usepackage{booktabs}
\usepackage{hyperref}
\usepackage{etoolbox}
\apptocmd{\thebibliography}{\setlength{\itemsep}{0.1pt}}{}{}

\newacro{TTS}{text-to-speech}
\newacro{CV}{computer vision}
\newacro{minDCF}{minimum detection cost function}
\newacro{EER}{equal error rate}
\newacro{SRE}{NIST speaker recognition evaluation}
\newacro{SV}{speaker verification}
\newacro{ASR}{automatic speech recognizer}
\newacro{DINO}{DIstillation with NO labels}
\newacro{LDA}{linear discriminant analysis}
\newacro{PLDA}{probabilistic \ac{LDA}}
\newacro{SSL}{self-supervised learning}
\newacro{AAM}{additive angular margin}
\newacro{MoCo}{momentum contrast}

\title{Non-Contrastive Self-Supervised Learning of Utterance-Level Speech Representations}

\name{Jaejin Cho$^{1}$, Raghavendra Pappagari$^{1}$, Piotr \.Zelasko$^{1,2}$, Laureano Moro-Velazquez$^{1}$, \\ \it Jes\'us Villalba$^{1,2}$, Najim Dehak$^{1,2}$}

\address{$^{1}$Center for Language and Speech Processing, Johns Hopkins University, Baltimore, MD, USA \\$^{2}$Human Language Technology Center of Excellence, Johns Hopkins University, Baltimore, MD, USA}

\email{\{jcho52,rpappag1,pzelasko,laureano,jvillal7,ndehak3\}@jhu.edu}

\begin{document}
\ninept
\maketitle
%JJ (TODO): Think about using this mane for the method: DINOSAUR (DINO in Speech Applications for Utterance-level Representation)
%
\begin{abstract}
%\vspace{-0.01in}
Considering the abundance of unlabeled speech data and the high labeling costs, unsupervised learning methods can be essential for better system development. One of the most successful methods is contrastive self-supervised methods, which require negative sampling: sampling alternative samples to contrast with the current sample (anchor). However, it is hard to ensure if all the negative samples belong to classes different from the anchor class without labels. This paper applies a non-contrastive self-supervised learning method on an unlabeled speech corpus to learn utterance-level embeddings. We used DIstillation with NO labels (DINO), proposed in computer vision, and adapted it to the speech domain. Unlike the contrastive methods, DINO does not require negative sampling. These embeddings were evaluated on speaker verification and emotion recognition. In speaker verification, the unsupervised DINO embedding with cosine scoring provided 4.38\% EER on the VoxCeleb1 test trial. This outperforms the best contrastive self-supervised method by 40\% relative in EER. An iterative pseudo-labeling training pipeline, not requiring speaker labels, further improved the EER to 1.89\%. In emotion recognition, the DINO embedding performed 60.87, 79.21, and 56.98\% in micro-f1 score on IEMOCAP, Crema-D, and MSP-Podcast, respectively. The results imply the generality of the DINO embedding to different speech applications.
% (TODO: JJ): I can do for analysis: clustering with different labels, embeddings from different position, compare embedding size}
% (Not fair comparison. Different configuration) In emotion recognition, the DINO embedding outperformed the x-vector embedding by 8.5, 4.7, and 8.4\% relatively in the micro-f1 score (\%) on IEMOCAP, Crema-D, and MSP-Podcast, respectively. 
\end{abstract}

\section{Introduction} \label{sec:data_description}
\Ac{SSL} is gaining more attention in many machine learning areas such as computer vision, natural language processing, and speech processing. \ac{SSL} does not require labeled data for model training. In many works, fine-tuned/post-processed \ac{SSL} models have shown promising results outperforming supervised methods when the same amount of labeled data is used~\cite{devlin-etal-2019-bert,lee2020biobert,NEURIPS2020wav2vec2,chen2020simclrv2,NEURIPS2020BYOL,caron2021dino}. 

In speaker verification, different self-supervised methods have been proposed as in~\cite{hsu2017unsupervised,stafylakis2019ss_spkemb,cho20tts_spkid,peng2020mixture,huh2020augmentation,xia2021moco_spkemb,zhang2021contrast_spkid}. 
%Some of the methods learn embedding by reconstructing a target sequence of acoustic frames~\cite{hsu2017unsupervised,stafylakis2019ss_spkemb,cho20tts_spkid,peng2020mixture}, while other methods utilize a contrastive loss or adapt the \ac{CV} techniques for speaker embedding learning~\cite{huh2020augmentation,xia2021moco_spkemb,zhang2021contrast_spkid}.
Some of these methods use a generative approach~\cite{hsu2017unsupervised,peng2020mixture,stafylakis2019ss_spkemb,cho20tts_spkid}, i.e., they learn to reconstruct the 
signal acoustic features from some latent representations. Usually, the goal is to
factorize the information into frame-/segment-level and utterance-level latent factors, expecting that the former will encode phonetic information while the latter will encode the speaker information. For example, the work in~\cite{cho20tts_spkid} used an architecture based on Tacotron 2 multi-speaker text-to-speech to learn speaker embeddings.

%methods reconstructing a target sequence of acoustic features aim to either factorize the representation, e.g., into frame-/segment-level and utterance-level latent factors~\cite{hsu2017unsupervised,peng2020mixture,stafylakis2019ss_spkemb,cho20tts_spkid} or factor out one information from the other, e.g., phonetic information from speaker information~\cite{stafylakis2019ss_spkemb,cho20tts_spkid}. However, they lack comparisons with large sizes of models and data.

\ac{SSL} methods based on contrastive loss are also popular~\cite{huh2020augmentation,xia2021moco_spkemb,zhang2021contrast_spkid} in speaker verification. Contrastive losses intend to make the current sample (anchor) close to the augmented version of the anchor (positive sample) while making the positive sample farther from the negative samples in their embedding space. In this context, negative samples are desired to be different semantically from the positive sample. Since the samples are unlabeled, most contrastive \ac{SSL} works in speaker verification compose negative samples just by randomly picking different samples to the anchor. However, in this random sampling, we are not sure if all the negative samples are from different classes w.r.t the positive sample. For example, when the anchor, thus also the positive sample, is an utterance from speaker A, there is a chance that some of the negative utterances come from speaker A as well. This could adversely affect the model training since the contrastive loss pushes the positive sample and negative sample farther to each other in the embedding space.

Non-contrastive methods, however, do not require negative samples, so they are free from the issue above. Moreover, non-contrastive methods have shown comparable or better performance compared to contrastive methods~\cite{NEURIPS2020BYOL,caron2021dino}. Considering these, we propose to apply a non-contrastive \ac{SSL} method originally proposed for~\ac{CV}, \ac{DINO}~\cite{caron2021dino}, that outperformed the previous \ac{SSL} methods in many \ac{CV} tasks, including linear and k-NN classification. 

Since \ac{DINO} training does not use explicit labels such as speaker or language IDs, we hypothesized the learned embedding to be general utterance-level embedding. In other words, the embedding may include attributes that are consistent within the utterance, such as speaker information, accent/language, emotion, and age. Thus, we evaluated the embedding not only for speaker verification but also for emotion recognition. The results confirmed that the \ac{DINO} embedding includes both speaker and emotion information.
%\section{Method}

%Distillation with no labels (DINO) is a non-contrastive method, first %proposed in \ac{CV} research. 
%As explained in the introduction, this method is free from the issue related to negative sampling during training. In this section, we explains the method more in detail and how it is applied to speech data.
\vspace{-0.05in}
\section{Distillation with NO labels in speech}
\vspace{-0.05in}
%Distillation with no labels (DINO) is a non-contrastive method, first proposed in \ac{CV} research. As explained in the introduction, this method is free from the issue related to negative sampling during training. In this section, we explains the method more in detail and how it is applied to speech data.
%
\begin{figure}[htp]
    \centering
    \includegraphics[width=1.0\linewidth]{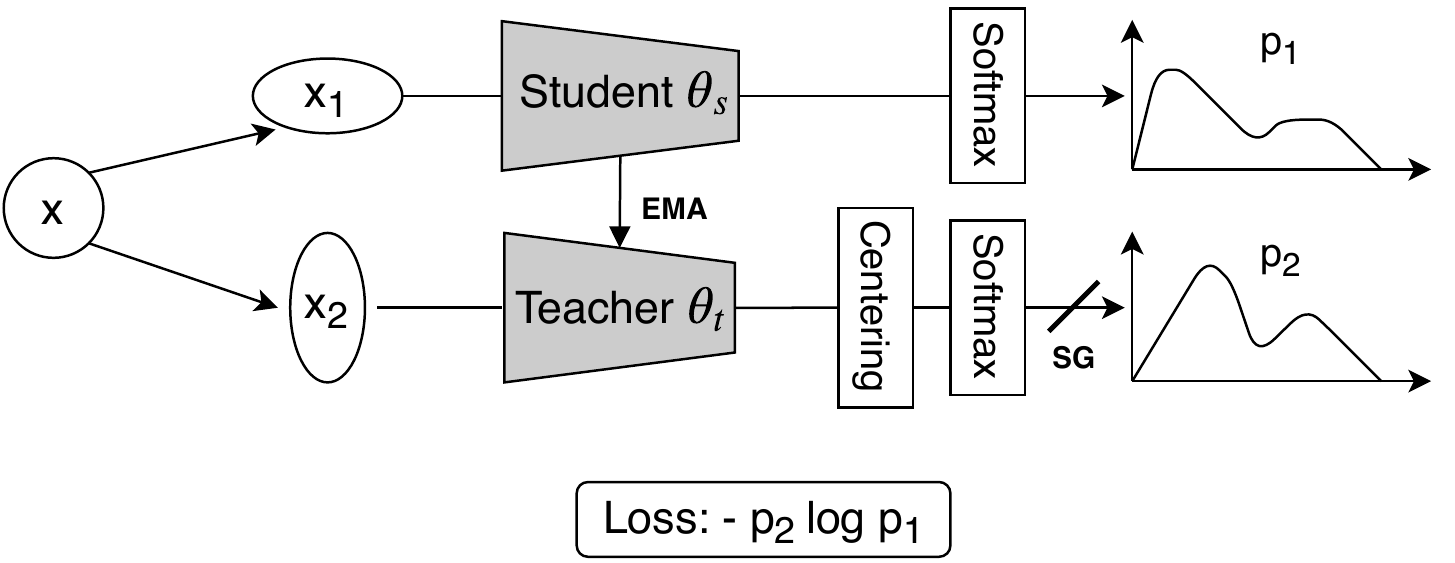}
    \caption{\ac{DINO} diagram. EMA stands for exponential moving average. SG stands for stop gradient. The figure shows a single augmented pair ($\textrm{x}_1$,$\textrm{x}_2$)  for simplicity.}
    \label{fig:dino}
    \vspace{-0.15in}
\end{figure}

In~\cite{caron2021dino}, the authors proposed a design to maximize the similarity between feature distributions of differently augmented images from an original image. This assumes that augmented images from one image keep the same semantic information. For example, although you crop two different portions from a dog image and make one a black and white image while making the other jittered, they are still dog images. 
This principle can be similarly applied to speech data. Assuming each utterance consists of a speech from one speaker, different segments extracted from the same utterance followed by noise augmentation share the same attributes that are consistent within the utterance, such as speaker information, accent/language, emotion, and age.
%The diagram for the method is in \todo{add}. Looking at the diagram from the bottom, %

DINO follows a knowledge distillation scheme, where the outputs of a teacher network 
are used as ground truth to optimize a student network. However, typical knowledge distillation uses a pre-trained teacher network, while DINO trains both networks in parallel. Fig.~\ref{fig:dino} depicts DINO's training scheme. % For example, in~\ac{CV}, a local view and a global view of an image are cropped where local and global views mean small and large portions of the image. Then, the cropped images are augmented with color jittering, Gaussian blur, solarization, etc. The local and global views are propagated through the student and teacher branch respectively to minimize the cross-entropy between two distributions where each distribution is calculated over the feature dimension. 
%First, we extract two different segments from a given utterance. Then, the extracted segments are augmented with sounds such as babbling, music, noise in the background or with room impulse response effects. The augmented segments, x1 and x2, are propagated through the student and teacher branch respectively to minimize the cross-entropy between two distributions, $\textrm{p}_1$ $\textrm{p}_2$ where each distribution is calculated over the feature dimension.
First, we augment a given utterance into a set of differently augmented segments, $S$. With speech data, for example, each segment is extracted randomly from the given utterance as either a short or long segment. Then, a sound such as babbling, music, noise in the background, or room impulse response effect is applied to each segment differently. The set $S$ includes two long segments, $s^{l}_1$ and $s^{l}_2$. All the augmented segments go through the student branch, while only the long segments go through the teacher branch. Each branch embeds information in the corresponding segments into the embedding vectors. Additionally, each network has a head with a softmax layer that classify each segment into a set of hypothetical classes. 
Thus, the student network $\theta_s$ is trained by minimizing,  
\begin{equation}
    \theta_s = \min_{\theta_s} \sum_{s \in \{s^{l}_1, s^{l}_2\}} \quad \sum_{\substack{s' \in S\\s'\neq \, s}} \quad H(p_2(s|\theta_t), p_1(s'|\theta_s)).
\label{eq:loss}
\end{equation}
where $H(a, b) = - a \cdot\log b$ is cross-entropy, and $p_1(.)$ and $p_2(.)$ are the softmax outputs of the student and teacher networks respectively. This loss intends to make the embeddings for all augmented versions of the utterance close between them, which relies on two assumptions. First, the long segments, used as input to the teacher, produce better representations than the short-segments. Second, the teacher network is always better than the student during the training, as explained below.
%\vspace{-0.1in}

The neural network architecture for student and teacher models comprises a backbone  followed by a projection head. %The backbone can be any encoder architecture such as ResNet~\cite{he2016resnet} or ViT~\cite{dosovitskiy2021vit}, and the projection head consists of 3 fully connected layers with their hidden dimension as 2048 followed by $l_\textrm{2}$ normalization and a weight normalized fully connected layer with $K$ dimensions.
The backbone can be any encoder that converts a sequence of vectors into a fixed-dimensional vector, e.g., using a pooling layer. The projection head consists of fully connected layers with non-linear activations. The student and teacher networks are initialized with the same architecture having the same parameters while they are updated in different ways during training. The student network is updated by gradient descent while the teacher network is updated by an exponential moving average on the student parameters, i.e., $ \theta_t \leftarrow \lambda \theta_t + (1-\lambda) \theta_s$. Parameter averaging is known to produce a better model~\cite{polyak1992acceleration,NIPS2017meanteacherbetter}, and this is also the case with the teacher network to be better than the student network during the training. The student model aims at the distribution from the teacher network to improve. To avoid a model to find trivial solutions, i.e., having distributions where one dimension is dominant or having uniform distributions, \textit{centering} and \textit{sharpening} are applied. \textit{centering} prevents one dimension from dominating by calculating a center and subtracting it from the logit before the softmax. However, \textit{centering} encourages a uniform distribution, and thus \textit{sharpening} is also applied where it encourages peaky distributions. This is done by setting a low value for the temperature in the teacher softmax normalization.
\vspace{-0.1in}
%\vspace{-0.05in}

\section{Experimental setup\protect\footnote{The code for experiments will be uploaded to the public repository \href{https://github.com/hyperion-ml/hyperion}{https://github.com/hyperion-ml/hyperion}.}} 

%\vspace{-0.1in}
\subsection{Encoder pre-training using DINO} 
\label{subsec:pretrain_DINO}
\vspace{-0.05in}
% (TODO???: what's different from DINO github)
% DINO architecture
We used a light ResNet34 (LResNet34) encoder from~\cite{villalba2018jhumitNISTsre2018} as the \ac{DINO} backbone considering available resources, with minor modifications: a kernel size of 3 in the first convolution layer, a mean and standard deviation pooling, and a following affine layer that outputs a 256-dimensional vector used as a embedding vector. The classification head consists of 3 linear layers with their hidden dimension as 2048, followed by $l_\textrm{2}$ normalization and a weight normalized linear layer with 65536 softmax output dimension, which were the best setup in the original \ac{DINO} paper~\cite{caron2021dino} that we found work well with speech data, too.

% Data augmentation
For the augmentation, we first extracted 4 short and 2 long segments randomly from a given utterance where we set 2 and 4 s for short and long segment length, respectively. We chose the specific numbers for the extracted segments considering our available computational resource and training time, but extracting more short segments could improve performance, as in~\cite{caron2021dino}. The extracted segments were augmented with babbling, music, noise in the background, and/or room impulse response effects. %The student network consumes all the 6 augmented segments while the teacher network only processes the 2 long segments.  
%\todo[inline]{JJ: Check the policy of data aug. w.r.t how it is applied if I have left space in the end}

% Training data description
VoxCeleb2 \textit{dev} set of 5,994 speakers was used for training DINO without speaker labels. VoxCeleb2 corpus~\cite{nagrani2017voxceleb1} consists of conversational speech utterances with moderate noise, which were processed from interview videos of 6,112 celebrities uploaded on Youtube, and it covers diverse ethnicity.

% applications
The final learned encoder was used for speaker verification and emotion recognition tasks employing other datasets.

\vspace{-0.1in} 
\subsection{Speaker verification} \label{sec:method_SV}
\vspace{-0.05in}
%\subsubsection{DINO embedding VS x-vector}
%\todo[inline]{JJ: fine-tuning experiments}
%In this experiment, we compared the \ac{DINO} embedding with x-vector. The same dataset and architecture as used in the \ac{DINO} training were used for x-vector except that x-vector used speaker labels for the training.

%We compared two different scoring back-ends: cosine similarity and LDA-PLDA. For the LDA-PLDA back-end training, we used the VoxCeleb1 \textit{dev} subset.
%\subsubsection{Speaker verification with no labels}
%\subsubsection{Iterative clustering and robust training}
% What we did for speaker verification: base model (DINO) -> iterative clustering -> robust training
In the speaker verification experiments, we followed a pipeline to build a progressively improving speaker verification system, which does not require speaker labels~\cite{thienpondt2020idlab}. Starting from the initial \ac{DINO} model trained in section \ref{subsec:pretrain_DINO}, the system development goes through the iterative clustering stage followed by the robust training stage. The training data was fixed as VoxCeleb2 \textit{dev} without speaker labels for the whole pipeline. %Through the series of experiments, we wanted to check 1) how the \ac{DINO} embedding performs compared to the best found \ac{SSL} embedding in the initial model training stage, 2) how the different initial models affect the later stages in the pipeline, and 3) how well the final system performs compared to the counterpart in supervised training.

In the iterative clustering stage, we trained a new larger model, ResNet34~\cite{he2016resnet} x-vector model, in a supervised way with the \ac{AAM}~\cite{deng2019arcface} loss based on pseudo speaker labels generated using the initial \ac{DINO} model. In detail, we extracted speaker embeddings from the initial model. Then, the embeddings were clustered using k-means clustering with 50k means, followed by agglomerative hierarchical clustering (AHC) with the number of clusters as 7500, which was heuristically determined for VoxCeleb2 \textit{dev}~\cite{thienpondt2020idlab}. The k-means clustering was used to make AHC computationally viable. Indices of the clusters are used as pseudo speaker labels for the supervised x-vector model training. The whole process was repeated 3 times until the speaker verification performance converged. During the process, the model parameters were continuously updated with the refined pseudo labels in each cycle. A 2-second segment was extracted per utterance to be used as a training sample.

In the robust training stage, we used a new larger model, Res2Net50~\cite{gao2019res2net} with 26 for the width of filters (in the first residual bottleneck block) and 4 for the scale hyper-parameter. The model was trained in a supervised way with pseudo labels generated from the ResNet34 model after 3 cycles of the iterative clustering stage. After the first 30 epochs of training, the post-pooling layers of the model were fine-tuned with a larger margin, 0.5, in the \ac{AAM} loss. A 2-second segment was extracted from each utterance to be used as a training sample, while 3-second segments were used in the large margin fine-tuning.

% trials that we evaluated our systems on
The learned embedding in each stage was evaluated for speaker verification on the original VoxCeleb1 \textit{test} (voxceleb1\_test\_o), VoxSRC-21 \textit{val}, or VoxSRC-21 \textit{test} trials. The latter two trials are from VoxCeleb Speaker Recognition Challenge2021(VoxSRC-21), where the challenge has a special focus on multi-lingual verification. Our team's submission of the system having 6.88 in EER(\%) in Table~\ref{tab:all_results} ranked third in track 3: self-supervised speaker verification where the participants were allowed to develop systems only with VoxCeleb2 \textit{dev} without using speaker labels.

\subsection{Emotion recognition}
% Most of the time, we can assume that there is only one dominant emotion in utterances shorter than 6s.
% The median duration of VoxCeleb2 \textit{dev} utterances is 6.08s.
The DINO model training uses segments (2s for $\textrm{x}_1$ and 4s for $\textrm{x}_2$ in Fig.~\ref{fig:dino}) drawn from VoxCeleb2 \textit{dev} dataset.
For the model to preserve emotion information, $\textrm{x}_1$ and $\textrm{x}_2$ have to share the same emotion at least for the majority of training.
As the median duration of utterances in VoxCeleb2 \textit{dev} is only 6.08s, we expect consistent emotion between $\textrm{x}_1$ and $\textrm{x}_2$ in general.
This assumption is also supported by the considered emotion datasets, which contain utterances of duration 2-11s with utterance-level emotion annotations. 
% annotation process  with only one emotion label), 
% it is reasonable to assume consistent emotion in VoxCeleb2 \textit{dev} utterances.
% these utterances, and hence we can assume that most of the time, the model receives segments of the same emotion per utterance.
Hence, we hypothesize that the model preserves emotion-related information along with speaker identity.

To probe the DINO embeddings for emotion information, we evaluated them for the emotion recognition task.
Specifically, we extracted DINO embeddings for three emotion recognition datasets, Crema-D~\cite{cao2014crema}, IEMOCAP~\cite{busso2008iemocap}, and MSP-Podcast~\footnote{Data provided by The University of Texas at Dallas through the Multi-modal Signal Processing Lab. This material is based upon work supported by the National Science Foundation under Grants No. IIS-1453781 and CNS-1823166. Any opinions, findings, and conclusions or recommendations expressed in this material are those of the author(s) and do not necessarily reflect the views of the National Science Foundation or the University of Texas at Dallas.}
 \cite{lotfian2017building}, and performed emotion classification using logistic regression on each corpus.
% Also, we evaluate the reusability of the DINO models for emotion recognition by adapting them for emotion recognition.
% For adaptation, we attach a fully-connected layer (with softmax activation) to the DINO model and train the entire model for emotion recognition.
% This process is usually referred to as fine-tuning in deep learning community.
% During fine-tuning process, the model is supposed to unlearn/discard speaker-related characteristics and focus on emotion characteristics.

The Crema-D dataset consists of acted emotions with utterances ranging from 2-4s in duration; The IEMOCAP dataset is composed of utterances with induced emotions and mostly 2-7s in duration; The MSP-Podcast dataset contains spontaneous utterances of duration 3-11s.
Regarding experimental setup, we performed leave-one-session-out cross-validation (5-fold CV) for IEMOCAP as in the previous works.
For Crema-D, we used the same train/dev/test splits as in~\cite{pappagari2020x}, and the standard splits for MSP-Podcast as in Release 1.4.
We used \textit{angry, happy, sad}, and \textit{neutral} emotions in the IEMOCAP and Crema-D dataset, and an additional emotion class \textit{disgust} in MSP-Podcast as in~\cite{pappagari2020x}.
We report weighted average values of class-wise f1-scores -- micro-f1 metric -- for these experiments.
For IEMOCAP, we report the average across five folds.
% As a baseline, we compare with~\cite{pappagari2020x} where authors used supervised x-vectors for emotion recognition.

% \todo{Do we need to include dataset details?}

%\vspace{-0.1in}
\section{Results}
\subsection{Speaker verification}
%\vspace{-0.3in}
%
\begin{table}[htbp]
\centering
\caption{Comparison between \ac{DINO}, \ac{MoCo}, and x-vector embeddings for speaker verification. The results are on the original VoxCeleb1 \textit{test} with \ac{EER}(\%) and MinDCF with $P_\mathrm{T}$=0.01. The PLDA back-end was trained with VoxCeleb1 \textit{dev} where its data size is 1/7 of VoxCeleb2 \textit{dev}.}
\vspace{-0.1in}
%\resizebox{\columnwidth}{!}{
\begin{tabular}{ccccc}%{p{0.9cm} p{1cm} p{1.1cm} p{1cm} p{1cm}}
\toprule
\multirow{2}{*}{}               & \multicolumn{2}{c}{Cosine scoring} & \multicolumn{2}{c}{PLDA} \\
                                & EER(\%)       & MinDCF       & EER(\%)          & MinDCF          \\ \hline \hline
\multicolumn{1}{r|}{\ac{DINO}} & 4.83 & 0.463 & 2.38 & 0.289       \\
\multicolumn{1}{r|}{\ac{MoCo}~\cite{thienpondt2020idlab}} & 7.3 & - & - & -        \\
\hline                             
\multicolumn{1}{r|}{x-vector} & 1.94 & 0.207 & 1.88 & 0.189         \\

%\multicolumn{1}{r|}{MoCo (w/o labels)} & -  & - & 7.3 & -      \\
\bottomrule

\end{tabular}
%}
\label{tab:cmpr_DINOnMoConxvec}
\end{table}
%\vspace{-0.1in}
%
%\vspace{-0.3in}
\subsubsection{DINO embedding in initial model training}
%\vspace{-0.05in}
In this experiment, we first compared two \ac{SSL} embeddings: \ac{DINO} and MoCo, in the initial model training stage in section~\ref{sec:method_SV}. This stage does not include iterative clustering and pseudo labeling for supervised training. As shown in Table~\ref{tab:cmpr_DINOnMoConxvec}, \ac{DINO} (non-contrastive method) outperforms \ac{MoCo} (contrastive method) by 40\% relatively in EER(\%). This implies that an \ac{SSL} model can learn embedding for speaker verification without negative sampling. 

\begin{table*}[htbp]
\centering
\caption{Speaker verification results over 3 different trial lists with progressing/different systems over the three stages. The numbers from~\cite{thienpondt2020idlab} seems rounded to the nearest tenth. Pseudo labels for robust training were generated from ResNet34 (iter3).}
\vspace{0.1in}
\begin{tabular}{|c|c|c|c|c|c|}
\hline
\multirow{2}{*}{Stage}                    & \multirow{2}{*}{Algorithm/Loss} & \multirow{2}{*}{Model}             & \multicolumn{3}{c|}{EER   (\%) with cosine scoring}         \\ \cline{4-6} 
                                          &                                 &                                    & voxceleb1\_test\_o & VoxSRC-21 \textit{val} & VoxSRC-21 \textit{test} \\ \hline
\multirow{2}{*}{\shortstack{Initial   model training \\ (self-supervised learning)}} & DINO                        & LResNet34                          & 4.83     & 13.96        &           -    \\ \cline{2-6} 
                                          & \ac{MoCo}                       & ECAPA~\cite{thienpondt2020idlab}                              & 7.3     &       -       &      -        \\ \hline
\multirow{5}{*}{Iterative clustering}     & \multirow{5}{*}{\shortstack{AAM loss \\ (margin=0.3)}}         & ResNet34 (iter1)                   & 2.56     & 8.59         &         -      \\ \cline{3-6} 
                                          &                                 & ResNet34 (iter2)                   & 2.13     & 7.35         &        -       \\ \cline{3-6} 
                                          &                                 & ResNet34 (iter3) & 2.13     & 6.97         &         -      \\ \cline{3-6} 
                                          &                                 & ResNet34 (iter4)                   & 2.14     & 6.88         &        -       \\ \cline{3-6} 
                                          &                                 & ECAPA (iter7)~\cite{thienpondt2020idlab}                      & 2.1     &    -          &        -       \\ \hline
Robust training                           & \multirow{2}{*}{\shortstack{AAM loss\\(margin=0.5)}}                          & \multirow{2}{*}{Res2Net50}                        & 1.89     & 6.50         & 6.88          \\ \cline{4-6}
+ larg-margin fine-tuning &                         &                          & 1.91     & 6.32         & 6.64          \\ \hline
\end{tabular}
%\vspace{-0.1in}
\label{tab:all_results}
\vspace{-5mm}
\end{table*}

Next, we compared the performance of \ac{DINO} to supervised x-vector in the DINO and  x-vector rows of Table~\ref{tab:cmpr_DINOnMoConxvec}. Their encoder architectures were identical (LResNet34). Although x-vector performed better than \ac{DINO}, it used VoxCeleb2-dev speaker labels while  \ac{DINO} with cosine scoring back-end did not use any labels at all. We also evaluated a PLDA back-end trained on VoxCeleb1-dev, which 7 times smaller than VoxCeleb. With PLDA where the gap between DINO and x-vector reduced further, while \ac{DINO} only used 1/8 of the labels than x-vector. The data used for PLDA also can be employed to fine-tune the DINO encoder, expecting further improvement as it is a common observation in most of the \ac{SSL} papers.

%The comparison between DINO and x-vector embeddings are shown in Table~\ref{tab:cmpr_DINOnMoConxvec}. \ac{DINO} with the cosine similarity back-end showed 4.83(\%) in EER which performed worse than the x-vector counterpart. However, the DINO speaker verification system did not use speaker labels at all. With the PLDA back-end, the DINO system improved further to 2.38 in EER(\%) %while it only used about 1/8 of the speaker labeled data used for the x-vector counterpart. 
%Comparing the DINO with PLDA to x-vector when the same amount of labeled data (vox1 \textit{dev}) was used, the DINO system outperformed the x-vector counterpart by large margin (2.38 vs 4.01 in EER(\%)).
%\vspace{-0.15in}
\subsubsection{Iterative clustering and robust training}
%\vspace{-0.05in}
The experimental results are shown in Table~\ref{tab:all_results} along the process from the \ac{DINO} initial model training to the robust training. In iterative clustering, the speaker verification performance is saturated around the 3rd iteration. This number of iterations until convergence is less than the one reported in~\cite{thienpondt2020idlab}\footnote{This is a rough comparison since detailed configurations are slightly different.}, possibly due to starting from a better initial model. Thus, we generated the pseudo labels from the model after the 3rd iteration (ResNet34 (iter3)) to use them for the last model training in the robust training stage, which improved further to 1.89 in EER(\%) on voxceleb1\_test\_o. This speaker verification system did not use any speaker labels in the development, and to compare, the supervised counterpart trained on about 2600 hours of speaker labeled data showed 0.93 in EER(\%). Finally, the large-margin fine-tuning did not improve on the voxceleb1\_test\_o trial list, while it improved on VoxSRC-21 \textit{val} and \textit{test} trial lists. %We observed that using increased length of segments, 4-second, for the large-margin fine-tuning degrades the performance on VoxSRC-21 \textit{test}. Thus, we used 3-second segments. We think this was due to the segments less than 4-second that take a large portion of the utterances in the \textit{test} trial pairs.

%\todo{JJ: Compare with Phonexia paper if I have left space in the end}
%\vspace{-0.1in}
\subsection{Emotion recognition}
%\vspace{-0.3in}
% \todo[inline]{JJ: Raghu will add more. I expect the amount to be around 1-page}
%

\begin{figure}
\centering
\includegraphics[width=0.40\textwidth]{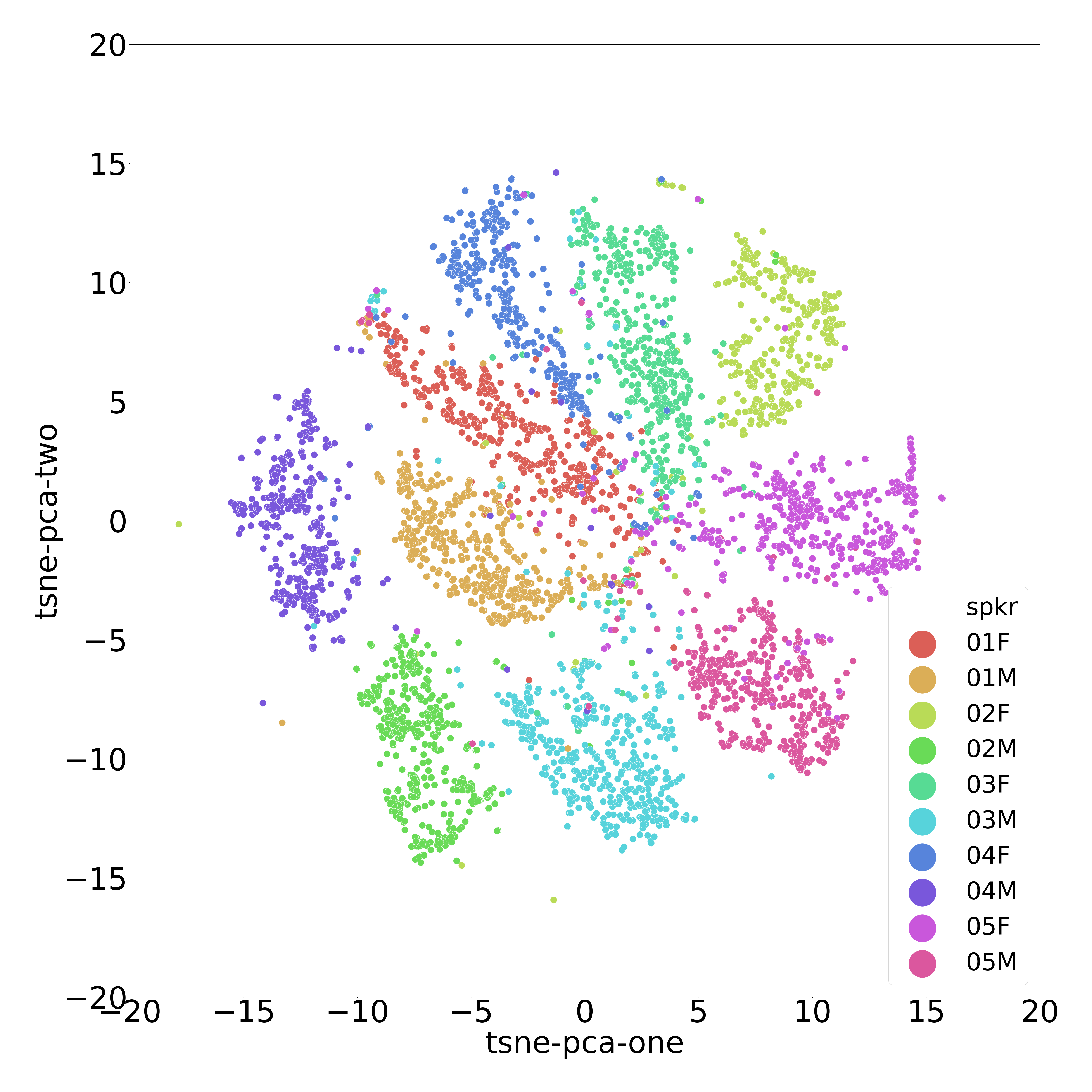}\label{subfig:IEMOCAP_DINOvectors_spkr}
\includegraphics[width=0.40\textwidth]{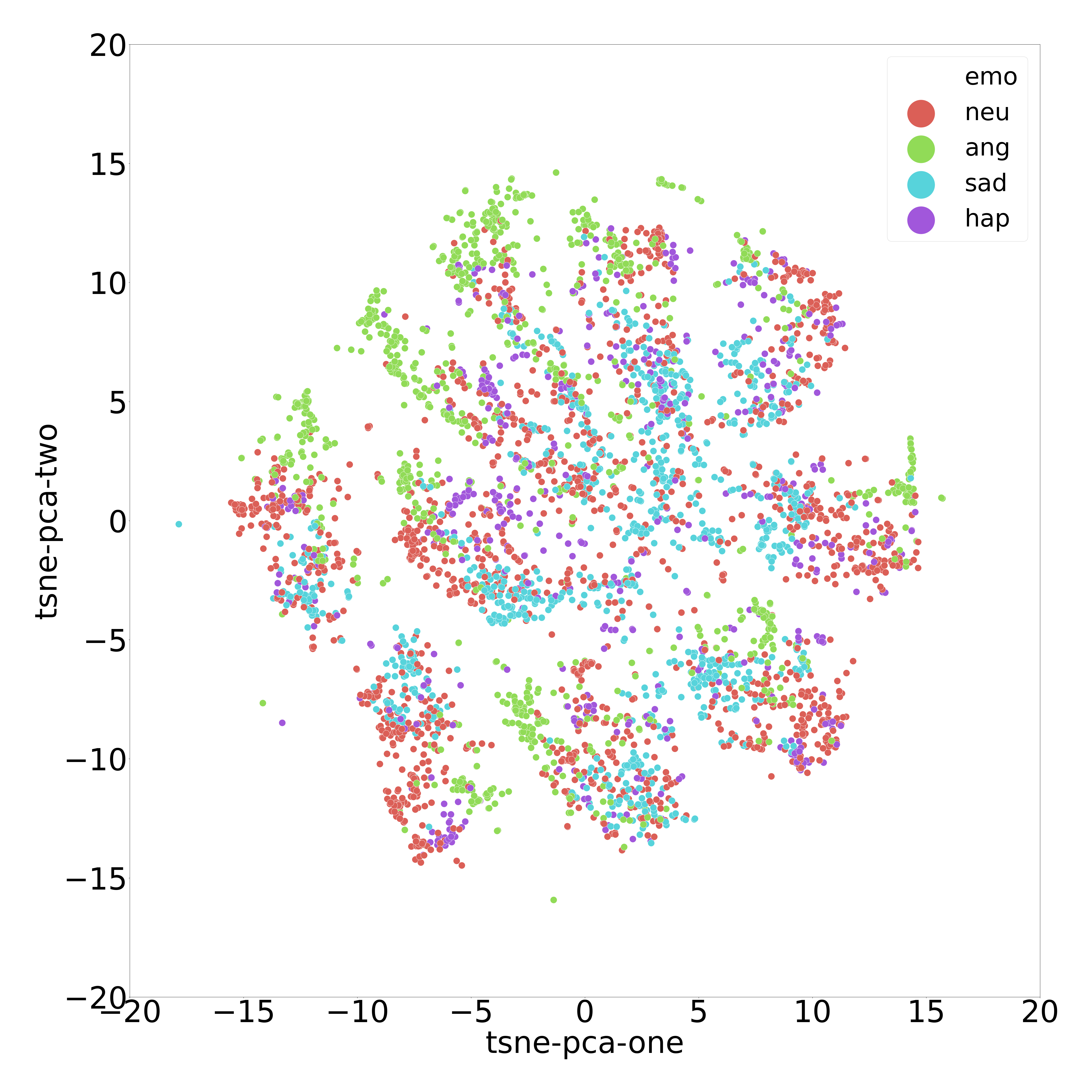}\label{subfig:IEMOCAP_DINOvectors_emo}
\caption{Analysis of DINO embedding space for IEMOCAP using t-SNE plots. Each color represents one speaker in the top plot and one emotion in the bottom plot.}
\label{fig:IEMOCAP_emb_space}
\end{figure}

\begin{table}[htbp]
\caption{Emotion classification results on three different dataset. All numbers in this table are micro-f1 (\%) scores}\label{tab:emotion_results}
\vspace{-0.1in}
\begin{tabular}{|l|l|l|l|}
\hline
         & IEMOCAP        & Crema-D        & MSP-Podcast    \\ \hline
x-vector~\cite{pappagari2020x} & 56.11          & 75.65          & 52.58          \\ \hline
DINO     & \textbf{60.87} & \textbf{79.21} & \textbf{56.98} \\ \hline
\end{tabular}
\vspace{-3mm}
\end{table}
%

%\vspace{-0.05in}
Table~\ref{tab:emotion_results} presents the results for emotion recognition using the DINO embedding.
We compare our results with~\cite{pappagari2020x}, where the authors trained a 
logistic regression classifier on top of
pre-trained x-vectors for emotion recognition. 
% This comparison is merely to provide expected range of the performance on these datasets as their supervised x-vectors are derived from a model that is different from our DINO model in several aspects.
Their x-vector model was pre-trained with 8kHz data to discriminate speakers. This includes telephone data other than VoxCeleb down-sampled to 8kHz where the final number of speakers was around 13k~\cite{villalba2019state}.
We observed that DINO embeddings performed better than the x-vectors suggesting that DINO embeddings contain more emotion predictive information than the x-vectors. 
This result makes sense since supervised x-vectors are trained to retain only speaker information and also trained with more data.
In contrast, the DINO model is trained to capture common information across extracted segments from the utterance.

Fig.~\ref{fig:IEMOCAP_emb_space} shows t-SNE plots of the DINO embedding space for the IEMOCAP dataset. 
From the top plot, we can observe clear clusters of speakers suggesting abundant speaker-relevant information in DINO embeddings.
In the bottom plot, there are signs of emotion clusters for some speakers, especially for angry and sad, suggesting some emotional information is captured in the representations.
Looking more closely within each speaker cluster, \textit{angry} and \textit{sad} are well separated which could be because they usually have distinct arousal levels (high for \textit{angry} and low for \textit{sad}).
% However, \textit{neutral} is overlapping with other emotions.

%\vspace{-0.1in}
\section{Conclusion}
%\vspace{-0.1in}
In this paper, we learned utterance-level embeddings based on \ac{DINO}, a non-contrastive self-supervised learning method originally proposed in~\ac{CV}. The embedding was evaluated on the speaker verification and emotion recognition tasks to check its generalizability. 
% speaker verification
DINO embedding achieved the state-of-the-art result in speaker verification when no speaker labels were used, outperforming the previous best contrastive \ac{SSL} embedding based on \ac{MoCo}. Also, it reduced the number of iterations until convergence in the iterative clustering stage.
% speech emotion recognition
When the \ac{DINO} embedding was used for emotion recognition, it performed better than the x-vector embedding that was found to contain emotion-related information in a previous study.
One thing to note is that the \ac{DINO} embedding was learned without any labels while it still performed competitively with the embedding using speaker labels for both speaker verification and emotion recognition. 
The \ac{DINO} method opens the way for leveraging unlabeled speech data, which is more easily available than labeled one.

% future work
Considering the \ac{DINO} embedding may also embed other attributes consistent within a given utterance, we will test the embedding on other speech applications such as accent/language, speech pathology, and age recognition. Also, fine-tuning the DINO embedding to one of the applications with the labeled data is a natural expansion of this work.

\section{Acknowledgements}
This project was supported by NSF Award 1816165.

\newpage
\bibliographystyle{IEEEtran}
\bibliography{mybib}

\begin{thebibliography}{10}
\providecommand{\url}[1]{#1}
\csname url@samestyle\endcsname
\providecommand{\newblock}{\relax}
\providecommand{\bibinfo}[2]{#2}
\providecommand{\BIBentrySTDinterwordspacing}{\spaceskip=0pt\relax}
\providecommand{\BIBentryALTinterwordstretchfactor}{4}
\providecommand{\BIBentryALTinterwordspacing}{\spaceskip=\fontdimen2\font plus
\BIBentryALTinterwordstretchfactor\fontdimen3\font minus
  \fontdimen4\font\relax}
\providecommand{\BIBforeignlanguage}[2]{{%
\expandafter\ifx\csname l@#1\endcsname\relax
\typeout{** WARNING: IEEEtran.bst: No hyphenation pattern has been}%
\typeout{** loaded for the language `#1'. Using the pattern for}%
\typeout{** the default language instead.}%
\else
\language=\csname l@#1\endcsname
\fi
#2}}
\providecommand{\BIBdecl}{\relax}
\BIBdecl

\bibitem{devlin-etal-2019-bert}
\BIBentryALTinterwordspacing
J.~Devlin, M.-W. Chang, K.~Lee, and K.~Toutanova, ``{BERT}: Pre-training of
  deep bidirectional transformers for language understanding,'' in
  \emph{NAACL}, Jun. 2019, pp. 4171--4186. [Online]. Available:
  \url{https://aclanthology.org/N19-1423}
\BIBentrySTDinterwordspacing

\bibitem{lee2020biobert}
J.~Lee, W.~Yoon, S.~Kim, D.~Kim, S.~Kim, C.~H. So, and J.~Kang, ``Biobert: a
  pre-trained biomedical language representation model for biomedical text
  mining,'' \emph{Bioinformatics}, vol.~36, no.~4, pp. 1234--1240, 2020.

\bibitem{NEURIPS2020wav2vec2}
\BIBentryALTinterwordspacing
A.~Baevski, Y.~Zhou, A.~Mohamed, and M.~Auli, ``wav2vec 2.0: A framework for
  self-supervised learning of speech representations,'' in \emph{NeurIPS},
  vol.~33.\hskip 1em plus 0.5em minus 0.4em\relax Curran Associates, Inc.,
  2020, pp. 12\,449--12\,460. [Online]. Available:
  \url{https://proceedings.neurips.cc/paper/2020/file/92d1e1eb1cd6f9fba3227870bb6d7f07-Paper.pdf}
\BIBentrySTDinterwordspacing

\bibitem{chen2020simclrv2}
T.~Chen, S.~Kornblith, K.~Swersky, M.~Norouzi, and G.~E. Hinton, ``Big
  self-supervised models are strong semi-supervised learners,'' \emph{NeurIPS},
  vol.~33, pp. 22\,243--22\,255, 2020.

\bibitem{NEURIPS2020BYOL}
J.-B. Grill, F.~Strub, F.~Altch\'{e}, C.~Tallec, P.~Richemond, E.~Buchatskaya,
  C.~Doersch, B.~Avila~Pires, Z.~Guo, M.~Gheshlaghi~Azar, B.~Piot,
  k.~kavukcuoglu, R.~Munos, and M.~Valko, ``Bootstrap your own latent - a new
  approach to self-supervised learning,'' in \emph{NeurIPS}, vol.~33, 2020, pp.
  21\,271--21\,284.

\bibitem{caron2021dino}
M.~Caron, H.~Touvron, I.~Misra, H.~J\'egou, J.~Mairal, P.~Bojanowski, and
  A.~Joulin, ``Emerging properties in self-supervised vision transformers,'' in
  \emph{ICCV}, 2021.

\bibitem{hsu2017unsupervised}
W.-N. Hsu, Y.~Zhang, and J.~Glass, ``Unsupervised learning of disentangled and
  interpretable representations from sequential data,'' in \emph{NeurIPS},
  2017, pp. 1876--1887.

\bibitem{stafylakis2019ss_spkemb}
\BIBentryALTinterwordspacing
T.~Stafylakis, A.~J. Rohdin, O.~Plchot, P.~Mizera, and L.~Burget,
  ``\BIBforeignlanguage{english}{Self-supervised speaker embeddings},'' in
  \emph{\BIBforeignlanguage{english}{Interspeech}}, vol. 2019, no.~9, 2019, pp.
  2863--2867. [Online]. Available:
  \url{https://www.fit.vut.cz/research/publication/12092}
\BIBentrySTDinterwordspacing

\bibitem{cho20tts_spkid}
J.~Cho, P.~Żelasko, J.~Villalba, S.~Watanabe, and N.~Dehak, ``{Learning
  Speaker Embedding from Text-to-Speech},'' in \emph{Interspeech}, 2020, pp.
  3256--3260.

\bibitem{peng2020mixture}
Z.~Peng, S.~Feng, and T.~Lee, ``Mixture factorized auto-encoder for
  unsupervised hierarchical deep factorization of speech signal,'' in
  \emph{ICASSP}, 2020, pp. 6774--6778.

\bibitem{huh2020augmentation}
J.~Huh, H.~S. Heo, J.~Kang, S.~Watanabe, and J.~S. Chung, ``Augmentation
  adversarial training for unsupervised speaker recognition,'' in \emph{NeurIPS
  Workshop}, 2020.

\bibitem{xia2021moco_spkemb}
W.~Xia, C.~Zhang, C.~Weng, M.~Yu, and D.~Yu, ``Self-supervised text-independent
  speaker verification using prototypical momentum contrastive learning,'' in
  \emph{ICASSP}, 2021, pp. 6723--6727.

\bibitem{zhang2021contrast_spkid}
H.~Zhang, Y.~Zou, and H.~Wang, ``Contrastive self-supervised learning for
  text-independent speaker verification,'' in \emph{ICASSP}, 2020, pp.
  6713--6717.

\bibitem{polyak1992acceleration}
B.~T. Polyak and A.~B. Juditsky, ``Acceleration of stochastic approximation by
  averaging,'' \emph{SIAM journal on control and optimization}, vol.~30, no.~4,
  pp. 838--855, 1992.

\bibitem{NIPS2017meanteacherbetter}
\BIBentryALTinterwordspacing
A.~Tarvainen and H.~Valpola, ``Mean teachers are better role models:
  Weight-averaged consistency targets improve semi-supervised deep learning
  results,'' in \emph{NeurIPS}, vol.~30, 2017. [Online]. Available:
  \url{https://proceedings.neurips.cc/paper/2017/file/68053af2923e00204c3ca7c6a3150cf7-Paper.pdf}
\BIBentrySTDinterwordspacing

\bibitem{villalba2018jhumitNISTsre2018}
J.~Villalba, N.~Chen, D.~Snyder, D.~Garcia-Romero, A.~McCree, G.~Sell,
  J.~Borgstrom, L.~P. Garc{\'i}a-Perera, F.~Richardson, R.~Dehak, P.~A.
  Torres-Carrasquillo, and N.~Dehak, ``State-of-the-art speaker recognition
  with neural network embeddings in nist sre18 and speakers in the wild
  evaluations,'' \emph{Comput. Speech Lang.}, vol.~60, 2020.

\bibitem{nagrani2017voxceleb1}
A.~Nagrani, J.~S. Chung, W.~Xie, and A.~Zisserman, ``Voxceleb: Large-scale
  speaker verification in the wild,'' \emph{Comput. Speech Lang.}, vol.~60,
  2020.

\bibitem{thienpondt2020idlab}
J.~Thienpondt, B.~Desplanques, and K.~Demuynck, ``The idlab voxceleb speaker
  recognition challenge 2020 system description,'' \emph{arXiv:2010.12468},
  2020.

\bibitem{he2016resnet}
K.~He, X.~Zhang, S.~Ren, and J.~Sun, ``Deep residual learning for image
  recognition,'' in \emph{Proceedings of the IEEE Conference on Computer Vision
  and Pattern Recognition (CVPR)}, June 2016.

\bibitem{deng2019arcface}
J.~Deng, J.~Guo, N.~Xue, and S.~Zafeiriou, ``Arcface: Additive angular margin
  loss for deep face recognition,'' in \emph{CVPR}, 2019, pp. 4690--4699.

\bibitem{gao2019res2net}
S.~Gao, M.-M. Cheng, K.~Zhao, X.-Y. Zhang, M.-H. Yang, and P.~H. Torr,
  ``Res2net: A new multi-scale backbone architecture,'' \emph{IEEE transactions
  on pattern analysis and machine intelligence}, 2019.

\bibitem{cao2014crema}
H.~Cao, D.~G. Cooper, M.~K. Keutmann, R.~C. Gur, A.~Nenkova, and R.~Verma,
  ``Crema-d: Crowd-sourced emotional multimodal actors dataset,'' \emph{IEEE
  transactions on affective computing}, vol.~5, no.~4, pp. 377--390, 2014.

\bibitem{busso2008iemocap}
C.~Busso, M.~Bulut, C.-C. Lee, A.~Kazemzadeh, E.~Mower, S.~Kim, J.~N. Chang,
  S.~Lee, and S.~S. Narayanan, ``Iemocap: Interactive emotional dyadic motion
  capture database,'' \emph{Language resources and evaluation}, vol.~42, no.~4,
  p. 335, 2008.

\bibitem{lotfian2017building}
R.~Lotfian and C.~Busso, ``Building naturalistic emotionally balanced speech
  corpus by retrieving emotional speech from existing podcast recordings,''
  \emph{IEEE Transactions on Affective Computing}, 2017.

\bibitem{pappagari2020x}
R.~Pappagari, T.~Wang, J.~Villalba, N.~Chen, and N.~Dehak, ``x-vectors meet
  emotions: A study on dependencies between emotion and speaker recognition,''
  in \emph{ICASSP}, 2020, pp. 7169--7173.

\bibitem{villalba2019state}
J.~Villalba, N.~Chen, D.~Snyder, D.~Garcia-Romero, A.~McCree, G.~Sell,
  J.~Borgstrom, F.~Richardson, S.~Shon, F.~Grondin \emph{et~al.},
  ``State-of-the-art speaker recognition for telephone and video speech: The
  jhu-mit submission for nist sre18.'' in \emph{Interspeech}, 2019, pp.
  1488--1492.

\end{thebibliography}

\end{document}